\documentclass[submission,copyright,creativecommons]{eptcs}
\providecommand{\event}{Types 2009} 
\usepackage[utf8]{inputenc}

\usepackage{graphicx}
\usepackage{mathpartir}
\usepackage{stmaryrd}
\usepackage{amsfonts}
\usepackage{array}
\usepackage{longtable}
\usepackage{bussproofs}
\usepackage{breakurl}

\newcommand{\unif}{\ensuremath{\stackrel{\scriptscriptstyle ?}{\equiv}}}

\newcommand{\MATITA}{Matita}

\newcommand{\N}{\ensuremath{\mathbb{N}}}
\newcommand{\Z}{\ensuremath{\mathbb{Z}}}
\newcommand{\Q}{\ensuremath{\mathbb{Q}}}
\newcommand{\GZ}{\ensuremath{\mathcal{Z}}}

\newcommand{\T}{\ensuremath{\mathbf{Type}}}
\renewcommand{\P}{\ensuremath{\mathbf{Prop}}}
\renewcommand{\L}{\ensuremath{\mathtt{List~}}}

\newcommand{\f}{\ensuremath{\mathtt{force}}}

\newcommand{\sG}{\ensuremath{\mathtt{SemiGroup}}}
\newcommand{\G}{\ensuremath{\mathtt{Group}}}
\newcommand{\aF}{\ensuremath{\mathtt{AssocFun}}}
\newcommand{\ac}{\ensuremath{\mathtt{assoc\_comp}}}
\newcommand{\one}{\ensuremath{\mathbf{1}}}
\newcommand{\carr}{\ensuremath{\mathtt{carr}}}
\newcommand{\op}{\ensuremath{\mathtt{op}}}
\newcommand{\inv}{\ensuremath{\mathtt{inv}}}
\newcommand{\unit}{\ensuremath{\mathtt{unit}}}
\newcommand{\zplus}{\ensuremath{\mathtt{zplus}}}

\newcommand{\match}{\ensuremath{\mathtt{match~}}}
\newcommand{\with}{\ensuremath{\mathtt{~with~[~}}}
\newcommand{\matchend}{\ensuremath{\mathtt{~]}}}

\newcommand{\ob}{\overrightarrow}

\newtheorem{example}{Example}

\title{Nonuniform Coercions via Unification Hints}
\author{Claudio Sacerdoti Coen
\institute{Department of Computer Science, University of Bologna}
 \email{sacerdot@cs.unibo.it}
 \and Enrico Tassi
 \institute{Microsoft Research-INRIA Joint Center}
 \email{enrico.tassi@inria.fr}}
\def\titlerunning{Nonuniform Coercions via Unification Hints}
\def\authorrunning{C. Sacerdoti Coen \& E. Tassi}

\begin{document}
\maketitle

\begin{abstract}
We introduce the notion of \emph{nonuniform} coercion, which is the
promotion of a value of one type to an enriched value of a different
type via a nonuniform procedure. Nonuniform coercions are a
generalization of the (uniform) coercions known in the literature and
they arise naturally when formalizing mathematics in an higher order
interactive theorem prover using convenient devices like canonical
structures, type classes or unification hints.  We also show how 
nonuniform coercions can be naturally implemented at the user level in an
interactive theorem prover that allows unification hints.
\end{abstract}

\section{Introduction}
In type theory, a coercion is a function $k$ of type $S \rightarrow T$
that can be automatically inserted by an interactive theorem prover to
promote a value $v$ of type $S$ to a value $k~v$ of type $T$ whenever
$v$ is used in a context where it is expected to have type $T$. The
typical example is the promotion of natural numbers to integers.

In a theory with dependent types coercions are better generalized to
couples $(k,n)$ where $k$ is a function of type $\Pi
\ob{x_i:S_i(x_1,\ldots,x_{i-1})}.T(x_1,\ldots,x_m)$ and $n \leq m$ is
the index of one of the arguments of $k$. The coercion is
automatically inserted to promote a value $v$ of type
$S(t_1,\ldots,t_{n-1})$ to a value
$k~t_1~\dots~t_{n-1}~v~t_{n+1}~\dots~t_m$ of type $T(t_1,\ldots,t_m)$.
The arguments $t_1,\ldots,t_{n-1},t_{n+1},\ldots,t_m$ are partially
inferred from the actual type of $v$ or from the expected type; those
that are not inferred become proof obligations for the user. The
typical example for the latter situation is the coercion from lists to
non-empty lists, that opens a new proof obligation for the
non-emptiness of the argument. This kind of parametric coercions have
been heavily exploited by Sozeau in the Russell language~\cite{russell-shortbib}.

All the previous examples of coercions are \emph{uniform} in the sense
that, up to the inferred arguments, the very same function $k$ is used
to promote the value $v$. For instance, if $k$ is the promotion from
natural numbers to integers, $3$ and $5$ are promoted respectively to
$k~3$ and $k~5$.  Nevertheless there are situations where we would
like to promote $v$ in a non uniform way, depending on the actual
value of $v$, but the language does not allow one to inspect (i.e. pattern
match over) $v$, only the meta-level allows it.

For example, consider the promotion of a type (the carrier of a
semi-group) to a semi-group (the carrier enriched with the operation).
Different types are 
associated (even not uniquely) to
different semi-groups on them: the type $\N$ of natural numbers could
be promoted to the semi-group $(\N,+)$ whereas the type $\L\N$ could
be promoted to $(\L\N,@)$ (where `$@$' is the append operation). Since
no function in type theory can distinguish between $\N$ and
$\L\N$, there is no uniform coercion of type $\T\to\sG$ 
that behaves in the expected way. 

As far as we know, nonuniform coercions have not been considered so
far in the literature.  Nevertheless, they arise quite naturally when
devices like canonical structures~\cite{canonical-structures}, type
classes~\cite{SozeauO08-shortbib} or unification hints~\cite{unification-hints}
are employed.

In Sect.~\ref{syntax} we introduce a syntax and semantics 
for nonuniform coercion
declarations and in Sect.~\ref{mathcomp} we present the use of 
nonuniform coercions as a complementary device to canonical structures
like mechanisms. In Sect.~\ref{uhintsnut} we recall the syntax and semantics 
of \emph{unification hints}~\cite{unification-hints}.
In Sect.~\ref{uhints} we show how nonuniform
coercions can be efficiently implemented at the user level in an
interactive theorem prover equipped with a flexible coercion system
and unification hints.
Sections~\ref{currymorph} and~\ref{notation} are devoted to the solution of
two different problems that naturally arise when nonuniform coercions
are employed.  Conclusions and future works follow in
Sect.~\ref{conclusions}.

\section{Syntax and semantics}
\label{syntax}
Let $\Gamma$ be a well-typed context, i.e. a sequence of variable declarations
of the form $x_i: R_i$ where each $R_i$ is a well typed type in the context
$x_1:R_1 \dots x_{i-1}:R_{i-1}$.
A nonuniform coercion declaration has the following syntax:

$$
\begin{array}{ccc}
 \begin{array}{lcl}
  \Gamma_1 & \vdash & 
     \begin{array}{lcl} 
         S_1 & \rightarrow & T_1 \\
         s_1 & \mapsto & t_1 
     \end{array}
 \end{array} &~~~~~ \ldots ~~~~~ &
 \begin{array}{lcl}
 \Gamma_n & \vdash & 
     \begin{array}{lcl} 
         S_n & \rightarrow & T_n \\
         s_n & \mapsto & t_n 
     \end{array}
 \end{array}
\end{array}
$$
where each $s_i$ has type $S_i$ in $\Gamma_i$; each $t_i$ has type $T_i$ in
$\Gamma_i$.

Promotion works as follows: let $\bar{s}$ be a term of type $\bar{S}$ which is
expected to have type $\bar{T}$ and let $\sigma$ be a substitution that
instantiates the variables declared in $\Gamma_i$ and such that
$s_i \sigma = \bar{s}$ and $S_i \sigma = \bar{S}$ and $T_i \sigma = \bar{T}$.
Then $\bar{s}$ can be promoted to $t_i \sigma$ of type $\bar{T}$.

Operationally, the substitution $\sigma$ is determined by looking for the
smallest index $i$ such that $\sigma$ is the most general unifier of
$s_i$ with $\bar{s}$, of $S_i$ with $\bar{S}$ and of $T_i$ with $\bar{T}$.
The substitution $\sigma$ can be made total by instantiating every
still uninstantiated variable $x_j:R_j\sigma$ with the result of a proof obligation for $R_j\sigma$.

Instead of stopping at the first index that yields a result, it could
also be possible to try all of them and let the user interactively choose the
desired promotion. Stopping at the first index is often desirable since it
allows one to add overlapping branches ordering them by number of open proof
obligations (see Example~\ref{ex3} below).

\begin{example}[Uniform coercions]
A uniform coercion $(k,n)$ where
$$
k:\Pi \ob{x_i:S_i(x_1,\ldots,x_{i-1})}.T(x_1,\ldots,x_m)
$$
is equivalent to a nonuniform coercion
$$
 \begin{array}{lcl}
  \ldots x_i:S_i(x_1,\ldots,x_{i-1}) \ldots & \vdash &
    \begin{array}{lcl}
    S_n(x_1,\ldots,x_{n-1}) & \rightarrow & T(x_1,\ldots,x_m) \\
    x_n & \mapsto & k~x_1~\dots~x_m 
    \end{array}
 \end{array}
$$
Conversely, every branch of a nonuniform coercion whose pattern $s_i$
is a variable $x_n$ is equivalent to a nonuniform coercion:
$$
 \begin{array}{lcl}
  \ldots x_i:S_i(x_1,\ldots,x_{i-1}) \ldots & \vdash &
    \begin{array}{lcl}
    S_n(x_1,\ldots,x_{n-1}) & \rightarrow & T(x_1,\ldots,x_m) \\
    x_n & \mapsto & t
    \end{array}
 \end{array}
$$
is equivalent to the uniform coercion $(\lambda \ob{x_i:S_i(x_1,\ldots,x_{i-1})}.t,n)$:
$$
\lambda \ob{x_i:S_i(x_1,\ldots,x_{i-1})}.t:\Pi \ob{x_i:S_i(x_1,\ldots,x_{i-1})}.T(x_1,\ldots,x_m)
$$
\end{example}
\vspace{0.3em}
\begin{example}[Structure enrichment] Let $\pi$ be a proof
of the associativity of $+$ over $\Z$ and $\bar \pi$ a proof of
the associativity of the append operation over lists.
$$
 \begin{array}{lcl}
   & \vdash & 
      \begin{array}{lcl}
        \T & \rightarrow & \sG \\
        \Z & \mapsto & (\Z,~+,~\pi) 
      \end{array} \\
 \end{array}
\quad\qquad
 \begin{array}{lcl}
   X: \T & \vdash & 
      \begin{array}{lcl}
        \T & \rightarrow & \sG \\
        \L X & \mapsto & (\L X,~@~X,~\bar\pi~X) 
      \end{array}
 \end{array}
$$
Note that the second coercion is polymorphic in the type
of the list arguments. Seen as two
independent uniform coercions, the two coercions are 
incoherent~\cite{coercivesubtyping}
since they both promote from and to the same couple of types and they
are not \mbox{$\alpha\beta\eta$-convertible}.
\end{example}
\vspace{0.3em}
\begin{example}[Property enrichment]\label{ex3}
Let $\pi$ be a proof that $+$ is associative, $\aF$ the structure
of associative functions and $\ac$ a proof that the composition of
objects in $\aF$ is in $\aF$.
$$
 \begin{array}{lcl}
    & \vdash & 
     \begin{array}{lcl}
       \N \to \N \to \N & \rightarrow & \aF~\N \\
       + & \mapsto & (+,\pi) 
     \end{array}\\
\\
     \begin{array}{l}
      X: \T, *: X \to X \to X,\\
       p: \forall a,b,c:X.a * (b * c)=(a * b) * c~ 
     \end{array}
     & \vdash & 
     \begin{array}{lcl}
      X \to X \to X & \rightarrow & \aF~X \\
      ~\!\!* & \mapsto & (*,p) 
     \end{array}\\
\\
   X: \T, f,g: \aF~X & \vdash &
      \begin{array}{lcl}
        X \to X \to X & \rightarrow & \aF~X \\
        f \circ g & \mapsto & \ac~f~g 
       \end{array}
 \end{array}
$$
When used to promote $+$ to an associative operation, no proof
obligation is left open. On the other hand, when used to promote $*:\N
\to \N \to \N$ to an associative operation, the user is left with the
proof obligation $p : \forall a,b,c:\N.a * (b * c)=(a * b) * c$.
Finally, the composition of two associative operations is promoted to
an associative operation applying the theorem $\ac$. The third case
can not be expressed as a uniform coercion since the pattern is not a
variable.
\end{example}

\section{Nonuniform coercions for mathematical structures}
\label{mathcomp}

Different ITPs provide the user different devices to fill the gap between
the high level language used to reason about abstract and complex mathematical
theories and the drastically lower level foundational language understandable
and checkable by a computer.  

Among the most widespread machineries to aid the user we mention
decision procedures; general purpose proof searching facilities, usually 
by means of external tools; model checkers and SMT solvers. All these tools
usually fall in the category called, with some abuse, automation. 
Even if these tools shown to be quite effective in many areas,
some radically different techniques were recently successfully 
employed~\cite{BigOps,canonical-structures,kacoq,type-classes-coq} 
to aid the user in systems based on higher order languages. 

Canonical structures, type classes (and their generalization into 
unification hints) allow the user to instrument the ITP linking 
objects and properties by means of structures.
Structures pack together axioms that give rise to a theory and that
are used to quantify theorems
(i.e. once defined the structure $\G$ packing together the
group axioms, theorems about groups have usually the shape 
$\forall G:\G, \ldots$).
The user is then asked to link models of these structures 
to their characterizing structure, so that the ITP is able to exploit 
such link when needed. 
For example, once proved that integers $\mathbb{Z}$ together with
addition $+$, $0$ and the inverse $-$ form a group $\GZ$, every
theorem that holds on groups can be used over expressions laying in the 
signature $(\mathbb{Z},+,-,0)$, even if the group structure $\GZ$ is not 
mentioned in any way in the context of the current conjecture. 
In addition to that, limited forms of Prolog like proof search allow the 
ITP to infer derived structures given the basic ones. 
For example, the pair $(0,0)$ can be transparently considered as the unit 
of the Cartesian product group $\GZ\times\GZ$, given
the general result that the Cartesian product of groups forms a group.

These forms of inference, although not as general 
as other automatic devices, tend to be more predictable 
and allow the user's formalization and proof strategy to be
closer to the widespread and natural mathematical argument:
\emph{since X together with Y form a Z, we have P(X,Y)}.
Moreover they shown to be very effective in building
reusable libraries of formalized
mathematics~\cite{canonical-structures}.

\subsection{Mathematical structures in dependently typed languages}
We briefly review here how the aforementioned devices are 
implemented in an interactive theorem prover based on an higher order
and dependently typed language.

Suppose that the system library contains the definition of natural numbers
$\N$, of multiplication over them and the proof that multiplication
is associative, commutative and it has a neutral element. Suppose also that
the library contains the definition of unital semigroups (i.e. semigroups
that have a left neutral element), represented as
a dependently typed record~\cite{pollackFAC02,coercions-record-shortbib}:
$$
\begin{array}{l}
\mathtt{UnitalSemiGroup} := \\
\quad\{ S: \T;~1:S;~*:S \to S \to S;~
\pi: \mathtt{is\_unital\_semi\_group~S~1~*}\}
\end{array}
$$
The record projection $S$ is declared as a uniform coercion of type
$$
\mathtt{UnitalSemiGroup} \to \T
$$ 
so that a quantification over a unital semigroup $G$ 
is automatically promoted to a quantification over the 
group carrier $(S~G)$.

The library contains all the interesting theorems of the theory of unital
semigroups, stated by making the quantifiers range over dependently typed
records. One example is the unicity of the neutral element, if it exists:
$$\forall G: \mathtt{UnitalSemiGroup}.\forall x:G.
  (\forall y:G. y*x=y) \Rightarrow x=1
$$
Suppose now that the user, during a proof over natural numbers, knows the
hypothesis $\forall y:\N. y+x=y$ (let's call it $H$) and she needs to
conclude that
$x=0$. In order to do that, she wants to apply the unicity property of the
neutral element of a unital semigroup, that is she wants to feed the
hypothesis $\forall y:\N. y+x=y$ to the theorem $L$ that proves the
unicity and that expects a unital semigroup $G$, an $x$ in the carrier of
$G$ and a proof that $\forall y:G. y+x=y$. The action of invoking the lemma $L$
generates the unification problem
$$\forall y:G. y*x=y \unif \forall y:\N. y+x=y$$
that can be solved by choosing for $G$ any unital semigroup over the natural
numbers whose operation is addition. The system automatically picks the
right group
definition if the user has already instructed the system
(by means of canonical structures or with the more flexible
mechanism of unification hints) to always enrich $\N$ to
$(\N,0,+,\ldots)$ under the previous constraints.

More formally, the user has fed the system with the request to apply the proof
term $L~?_G~x~H$ where $?_G$ is a metavariable to be instantiated by unification
and the unifier instantiates $?_G$ with $(\N,0,+,\ldots)$. Note that no
coercion has been applied in this case: $?_G$ is identified by the system
since it occurs in the type of $x$ and $H$, which are known.

Most of the time, the previous mechanism is sufficient to perform the
enrichment. Nevertheless, there are situations where the enriched structure
to be discovered does not occur dependently in the type of some other parameter
(or in the expected conclusion). 
For instance, suppose that we already know that the exponential function is
injective and suppose that we want to inhabit the type
$(\mathtt{InjectiveFun}~\mathbb{R}~\mathbb{R})$ of injective functions over
real numbers with the function $\lambda x.e^{e^x}$. Since the previous
$\lambda$-abstraction is just a type theoretic function from $\mathbb{R}$ to
$\mathbb{R}$, what we need here is an automatic promotion of the function
from $(\mathbb{R} \to \mathbb{R})$ to
$(\mathtt{InjectiveFun}~\mathbb{R}~\mathbb{R})$, which can be achieved
by means of a nonuniform coercion similar to the one in Example~\ref{ex3}
that states that the composition of injective functions is injective.

Without nonuniform coercions, the user is obliged to manually feed the
system with the enriched structure itself, as in a traditional procedural
language. 
This is, for instance, what happens currently in~\cite{canonical-structures},
where two distinct mechanisms (and notations) are used to enrich structure
when the canonical structure mechanism is not triggered.
The simplest one overloads every notation defining a scope for
every possible enrichment. Then, the explicit scope delimiter can be used
to interpret the notation as desired by the user. For example the 
construction $A\cap B$,
even if $A$ and $B$ are groups, returns a set, while $(A\cap B)\%G$ 
returns a group.
This mechanism does not only force to redefine every notation for every
structure, but has also some technical limitations that make it fail if
one among $A$ and $B$ is not explicitly typed as a 
group.\footnote{This 
phenomenon happens, for example, when $B$ is a local definition for $C\cap D$.
Unfolding $B$ and then forcing the groups scope would make all instances of
$\cap$ to be interpreted at the group level, but if $B$ is not manually
unfolded, only one occurrence of $\cap$ is affected by the scope delimiter,
thus the full expression fails to be enriched.}
The second mechanism is way more verbose,
but does not suffer from the aforementioned limitations, and is thus
used as a fall back for the former. To 
enrich G to the wanted $\mathtt{structure}$ the user types
``[the $\mathtt{structure}$ of G]'' and a rather complex machinery
generates behind the scenes an ad-hoc unification problem that triggers
canonical structures inference.\\ 
Note that in both approaches what the users types is what will then be
displayed, thus the more verbose an enrichment notation is, the more
cluttered the lemmas statements and the intermediate goals of their proofs 
will be.

In the next section we see that,
quite surprisingly, nonuniform coercions can be implemented for free at the
user level in a system based on dependent types and unification hints.

\section{Unification hints in a nutshell}
\label{uhintsnut}

Unification hints~\cite{unification-hints} give solutions to (higher order)
unification problems that fall outside the domain of the regular unification
heuristic implemented by the ITP. They are presented as rules of the following
form:
\begin{prooftree}
\AxiomC{$\ob{?_x}\;:=\;\ob{H}$}
\LeftLabel{$\Gamma \vdash$}
\RightLabel{myhint}
\UnaryInfC{ $ P\;\equiv\;Q$ }
\end{prooftree}
where:
\begin{enumerate}
 \item $\Gamma$ is a context that declares meta-variables $?_y$;
 \item $\ob{?_x}\;:=\;\ob{H}$ is a telescope of definitions for
       metavariables undeclared in $\Gamma$;
 \item all meta-variables in $P$ and $Q$ are declared in $\Gamma$ or
       defined in the telescope;
 \item $P \equiv Q$ is a {\em linear} pattern in the meta-variables
       defined in the telescope (i.e. every
       meta-variable $?_x$ occurs just once either in $P$ or in $Q$);
 \item the type checking rules for meta-variables mimick the corresponding
       rules for variables; all the terms in $\Gamma$,
       $\ob{H}$, $P$ and $Q$ must be well typed. We use meta-variables since
       we expect them to be instantiated by unification\footnote{
Due to type dependencies, it is not always possible to give all the
meta-variable declarations first (in $\Gamma$) and then all the meta-variable
definitions. Interleaving declarations and definitions
poses no problem (and we implicitly assume that it is done). Nevertheless,
we prefer to present the hints in this way for clarity purposes.
}.
\end{enumerate}

A hint is {\em acceptable} if
$\Gamma \vdash P[\ob{H}/\ob{?_x}] \equiv Q[\ob{H}/\ob{?_x}]$,
i.e. if the two  terms obtained by telescopic substitution,
are {\em convertible} in $\Gamma$.
Since convertibility
is (typically) a decidable relation, the system is able to discriminate
acceptable hints. As a consequence, for every substitution $\sigma$ that
instantiates only meta-variables in $\Gamma$, all unification problems of the
form $P \sigma \unif Q \sigma$ can be solved by the unifier $\rho$ that assigns
to each $?_x$ the term $H\sigma$. As a generalization,
if $\tau$ is a generic meta-variable substitution,
all unification problems of the form $P \tau \unif Q \tau$
can be solved by recursively solving the new problems
$\ob{?_x \tau \unif H \tau }$ (whose solution is $\rho$ in the
simple case where $\tau$ behaves as the identity function on the meta-variables
in the telescope).

Unification hints are triggered in case of a unification failure. If the
failing problem is $\bar P \unif \bar Q$, the system looks for an
hint such that there exists a unifier $\tau$ of $P$ with $\bar P$ and
$Q$ with $\bar Q$, and then try to solve the unification problem by triggering
the recursive problems $\ob{?_x \tau \unif H \tau }$.

When there are multiple hints matching the failing unification problem, the
system can either ask the user about the desired one, or it can
pick the first matching hint according to some user defined precedence
level (e.g. in order of declaration or by more precise matching).

We give a bird's-eye view on how they allow one to apply a simple property of
groups in a context where the group structure is implicit. Consider the
following statement and its corresponding form without (part of its) notational
sugar.
$$
a + 0 = a \qquad \zplus~a~0 = a
$$
\noindent
where $a$ is a point in $\Z$. The group theoretical property
we want to apply is the right-identity law for $+$ and $0$ that
is
$$
\mathtt{grid} : \forall G:\G, 
        \forall x: \carr~G, \op~G~x~(\unit~G) = x 
$$
Instantiating this axiom to $a$ and leaving $G$ implicit (i.e. considering the
term $\mathtt{grid}~?_G~a$) triggers the unification problem 
$\carr~?_G \unif \Z$, since $a:\Z$ is expected to have type
$\carr~?_G$. This problem can be solved only 
by guessing a model of a group whose carrier is the set $\Z$.

Similarly, if we use $\mathtt{grid}$ to perform the rewriting without
instantiating it first (i.e. we use the term $\mathtt{grid}~?_G~?_x$ with
an expected type whose left hand side is $\zplus~a~0$), we trigger
the unification problem:

$$\op~?_G~?_x~(\unit~?_G) \unif \zplus~a~0$$

If the unification goes from left to right, the system must unify
$\op~?_G$ with $\zplus$; if it goes from right to left, it must
unify $\unit~?_G$ with $0$. In both cases we are facing again an
unification problem where a projection ($\op,\unit$ or $\carr$) is applied to an implicit structure $?_G$ and a model of the structure must
be guessed by the system. Since any guess would be arbitrary and could involve
proof search, we expect the system to fail.

To force a solution to the unification problem, the user can
specify the following hints that link the carrier $\Z$, the operation
$\zplus$ and the constant $0$ to the group structure $\GZ$.

\begin{center}
\AxiomC{$?_g := \GZ $}
\LeftLabel{$\vdash$}
\UnaryInfC{ $ \carr~?_g \equiv \Z$ }
\DisplayProof
$\quad$
\AxiomC{$?_g := \GZ $}
\LeftLabel{$\vdash$}
\UnaryInfC{ $ \op~?_g \equiv \zplus$ }
\DisplayProof
$\quad$
\AxiomC{$?_g := \GZ $}
\LeftLabel{$\vdash$}
\UnaryInfC{ $ \unit~?_g \equiv 0$ }
\DisplayProof
\end{center}

Unification hints can also be used to drive proof search with clauses that
resemble the ones used in logic programming. For instance, to solve the
problem
$$
\carr~?_1 \unif \Z \times \Z
$$
the user can declare the following hint that recursively reduces the problem
to simpler ones:
\begin{prooftree}
\AxiomC{$?_A := \carr~?_h$}
\AxiomC{$?_B := \carr~?_q$}
\AxiomC{$?_g := ?_h \times ?_q$}
\LeftLabel{$?_h,?_q: \G \vdash$}
\TrinaryInfC{
  $ \carr~?_g \equiv ?_A \times ?_B$ }
\end{prooftree}
Intuitively, the hint says that, if
the carrier of a group $?_g$ is a product $?_A \times ?_B $, 
where $?_A$ is the carrier of a group $?_h$ and $?_B$ is the carrier 
of a group $?_q$ then a solution consists in choosing for $?_g$
the group product of $?_h$ and $?_q$.

\subsection{Hints and coercions indexing}

Another view at unification hints is that they define equivalence classes of
convertible terms that are considered indistinguishable by every functionality
of the system that works up to unification. Promotion of terms via (uniform)
coercions is the typical example: if the user declares a coercion from
$\Z$ to $\Q$, we expect the system to be able to promote also
$\carr~\GZ$ to $\Q$.

This behaviour, however, does not come for free. Consider a failing unification
problem $\bar P \unif \bar Q$. In order to retrieve a coercion from $\bar P$
to $\bar Q$, it
would be inefficient to iterate over the whole set of coercions to find the
ones that goes from $M$ to $N$ such that $M$ unifies with $\bar P$ and $N$
unifies with $\bar Q$. The usual implementation strategy is to use a
discrimination tree~\cite{McCune,nieuwenhuis01evaluation-shortbib} 
(or discrimination nets) to index the coercion and perform a quick
approximated search. These data structures, born in the field of
automatic theorem proving, are first order oriented and compare terms
according to their rigid structure. For example 
$(\mbox{carr }\GZ)$ and $\Z$ would be considered different for at
least two reasons: both their head constant and head function symbol arity
differ.

Thus, in order to retrieve coercions up to hints, the
system must do some additional work, for instance to index every coercion
multiple times (on all $M'$ and $N'$ such that $M$ and $M'$ are in the same
hint-induced equivalence class, and the same for $N$ and $N'$), or to perform
the search multiple times (for every $P'$ and $Q'$ such that $\bar P$ and
$P'$ are in the same hint-induced equivalence class, and the same for
$\bar Q$ and $Q'$).

In the rest of the paper we assume our system to implement both unification
hints and retrieval of coercions up to unification hints. The
\MATITA{} interactive
theorem prover~\cite{ck-sadhana-shortbib,matita-jar-uitp-shortbib} satisfies these requirements.

\section{Non Uniform Coercions via Unification Hints}
\label{uhints}

We analyze at first a particular scenario where it is easier to explain the
ideas involved.
Suppose that we are interested in implementing the following nonuniform
coercion with two branches:
$$
\begin{array}{ccc}
 \begin{array}{lcl}
  & \vdash & 
  \begin{array}{lcl} 
      S & \rightarrow & T \\
      s_1 & \mapsto & t_1 
  \end{array}
 \end{array} & ~~~ &
 \begin{array}{lcl}
  & \vdash & 
  \begin{array}{lcl} 
      S & \rightarrow & T \\
      s_2 & \mapsto & t_2 
  \end{array}
 \end{array}
\end{array}
$$
The simplified scenario is characterized by the fact that all branches are
uniform in the types $S$ and $T$; the fact that we consider just two branches
and that we analyze only the case where all contexts $\Gamma_i$ are empty
is just to simplify the presentation and dropping these two limitations poses
no real problem.

Suppose also, to grant the coherence of the coercion graph, 
that the system only allows one
to declare a single uniform coercion between two given
(equivalence classes of) types. The only way in which a single uniform coercion
can present the expected non uniform behaviour is that it takes $t_i$ in input:
$$
\begin{array}{l}
k : \forall S:\T.S \rightarrow \forall T:\T. T \to T\\
k~S~s~T~t := t
\end{array}
$$
so that $s_1$ can be promoted to $k~S~s_1~T~t_1$ and $s_2$ can be promoted to
$k~S~s_2~T~t_2$. However, when a term $\bar s$ of type $S$ is expected to have
type $T$, according to the standard semantics of uniform coercions, the user
will be presented with a proof obligation of type $T$ and the term $\bar s$ will
be promoted to $k~S~s~T~t$ (that reduces to $t$) where $t$ is the proof term
provided by the user in the proof obligation. This means that the coercion
maps any term of type $S$ to a term of type $T$ after asking the user what term
must be chosen. Instead, we would like the system to automatically pick $t_1$ for $t$
when $\bar s$ is $s_1$, to pick $t_2$ for $t$ when $\bar s$ is $s_2$ and to
fail in all the other cases, without bothering the user at all.

The idea to achieve the expected result is to use unification hints to
automatically suggest the term $t$. However, unification hints are only
triggered in case of a unification failure and apparently the only
failing unification problem is the initial one: $S \unif T$ where $s$ and $t$
do not occur at all.

\subsection{Lifting terms to types}

In a dependently typed language, terms can occur in types, or better
can be lifted to types when they occur as arguments of a dependently 
typed function. 

The trick we employ is to declare the nonuniform coercion not from $S$ to $T$,
but from $S$ to a type that is convertible to $T$ but that exposes $s$ and $t$.
We can make a first shot with the following definitions:
$$
\begin{array}{l}
\f : 
  \forall S:\T. S \rightarrow \forall T:\T. T \rightarrow \T\\
\f~S~s~T~t := T
\end{array}
$$
\noindent
The $\f$ type has the property that
$\f~S~s~T~t$ is $\beta$-equivalent to $T$. The coercion $k$ is
now redefined as follows:
$$
\begin{array}{l}
k : \forall S:\T.\forall s:S. \forall T:\T.\forall t:T.
 \f~S~s~T~t\\
k~S~s~T~t := t
\end{array}
$$
With these definitions, when a term $\bar s$ of type $S$ is used with type $T$,
the system tries to promote $\bar s$ to $k~S~\bar s~?_T~?_t$ (which reduces to $?_t$)
yielding a new unification problem
$$\f~S~\bar s~?_T~?_t \unif T$$
In this unification problem we have both $\bar s$ and $?_t$, and so we
have the data for choosing a term for $?_t$ in function of $\bar s$. 
Moreover, since $?_t$ is also the argument the coercion $k$ is returning,
the solution for the unification problem can define the output of the 
coercion.

However,
we also have that the left hand side reduces to $?_T$ and thus the system
can easily solve the unification problem by choosing $T$ for $?_T$, without
using any unification hint at all and, once again, opening a proof obligation
of type $T$ to determine the unconstrained term $?_t$.

\subsection{Locked reduction and lockpicking via hints}

In order to solve the problem, we need to change our definition of $k$
(and $\f$) again in order to produce in a similar way a 
new unification problem whose solution is too difficult to be found 
by the system without resorting to unification hints.

In other words, we need to change the definition of $\f$ in such
a way that $(\f~S~\bar s~?_T~?_t)$ is no longer always reducible to
$?_T$, but only in certain user-defined situations.

The latter behaviour can be achieved by adding an additional parameter to
$\f$ that must take a precise value to unlock the good reduction.
This can be achieved in many ways, one of them being to parameterize
$\f$ over an inhabitant of the unit type and using pattern matching
on it to block reduction until the canonical inhabitant is passed:

$$
\begin{array}{l}
\f : 
  \forall S:\T. S \rightarrow \forall T:\T. T \rightarrow \one \rightarrow  \T\\
\f~S~s~T~t~lock := \match lock \with \star \Rightarrow T \matchend
\end{array}
$$
\noindent
The $\f$ type has the property that
$(\f~S~s~T~t~\star)$ is $\beta$-equivalent to $T$ (where $\star$ is
the canonical inhabitant of $\one$), but
reduction of $(\f~S~s~T~t~?_l)$ is blocked.
The coercion $k$ is now redefined as follows:
$$
\begin{array}{l}
k : \forall S:\T.\forall s:S. \forall T:\T.\forall t:T.
 \one \rightarrow \f~S~s~T~t\\
k~S~s~T~t~lock := \match lock \with \star \Rightarrow t \matchend
\end{array}
$$
The system now tries to promote $\bar s$ of type $S$ to
$(k~S~s~?_T~?_t~?_l)$ that generates the new failing unification problem
$$\f~S~\bar s~?_T~?_t~?_l \unif T$$
At last, this is where the unification hints come into play. Indeed, we can
define the following two acceptable unification hints:

\begin{center}
\AxiomC{ $     ?_T := T$}
\noLine
\UnaryInfC{ $     ?_t := t_1$}
\noLine
\UnaryInfC{ $     ?_l := \star$}
\LeftLabel{$\vdash$}
\UnaryInfC{$\f~S~s_1~?_T~?_t~?_l~\equiv~T$}
\DisplayProof
$\quad$
\AxiomC{$     ?_T := T$}
\noLine
\UnaryInfC{ $     ?_t := t_2$}
\noLine
\UnaryInfC{ $     ?_l := \star$}
\LeftLabel{$\vdash$}
\UnaryInfC{$\f~S~s_2~?_T~?_t~?_l~\equiv~T$}
\DisplayProof
\end{center}

\noindent
that suggests respectively the solution $t_1$ when $s_1$ is matched and
the solution $t_2$ when $s_2$ is, unblocking the reduction by fixing $?_l$ to
be $\star$ so that $(\f~S~s_i~T~t_i~\star)$ reduces to $T$ and
the promoted term $(k~S~s_i~T~t_i~\star)$ reduces to $t_i$ as expected.

\subsection{The general solution}
The simplest scenario of the previous section assumed every branch of the
nonuniform coercion to go from $S$ to $T$. The assumption was necessary to
type the very first failing attempts. However, it is useless for the final
working solution presented at the end of the section and thus it can be dropped.
Similarly, the number of branches in the nonuniform coercion also plays no
role since it just corresponds to the number of unification hints to be
declared. Also the order in which the branches are listed (and that determines
the branch that is triggered when multiple branches can) can be preserved by
defining the unification hints in the good order (or by giving an explicit
precedence to them). Finally, branches with a non empty context $\Gamma_i$ also
pose no problem since unification hints are also parameterized by a context.

Thus, to summarize, the final general solution is the following. First of all,
we declare at the beginning of the library the definition of $\f$
and $k$
$$
\begin{array}{l}
\f : 
  \forall S:\T. S \rightarrow \forall T:\T. T \rightarrow \one \rightarrow  \T\\
\f~S~s~T~t~lock := \match lock \with \star \Rightarrow T \matchend\\\\
k : \forall S:\T.\forall s:S. \forall T:\T.\forall t:T.
 \one \rightarrow \f~S~s~T~t\\
k~S~s~T~t~lock := \match lock \with \star \Rightarrow t \matchend
\end{array}
$$
and we declare $k$ as the only (uniform) coercion.
Then, to declare a nonuniform coercion whose branches are all of the form
$$
 \begin{array}{lcl}
  \Gamma_i & \vdash & 
     \begin{array}{lcl} 
         S_i & \rightarrow & T_i \\
         s_i & \mapsto & t_i 
     \end{array}
 \end{array}
$$
the user declares for each branch the following acceptable unification hint:
\begin{prooftree}
\AxiomC{$?_T := T'_i$}
\AxiomC{$?_t := t'_i$}
\AxiomC{$?_l := \star$}
\LeftLabel{$\Gamma'_i \vdash$}
\TrinaryInfC{$\f~S'_i~s'_i~?_T~?_t~?_l~\equiv~T'_i$}
\end{prooftree}
\noindent
where $\Gamma'_i$ is obtained from $\Gamma_i$ by turning every variable
$x$ in a meta-variable $?_x$, and the same holds for
$T_i'$, $S'_i$ and $s'_i$ and $t'_i$.

Of course, the system could just implement the standard syntax for nonuniform
coercions as syntactic sugar for the unification hint itself. 

In particular, since only one uniform coercion ($k$) is declared, the ITP
code that implements coercion can be simplified, for example dropping
all optimizations for fast indexing/retrieval. Actually, since the coercion
$k$ must be able to promote any type to $\f$, the ITP must allow the declaration
of coercions going from any type to something. This feature can be considered
quite unusual since it does not allow any form of efficient indexing. For
instance, the Coq system would not accept $k$ as a coercion. \MATITA{}, instead,
does not have this restriction. In any case, with a bit of work, we can avoid
the limitation when it is there by redeclaring $k$ multiple times as a coercion
from $T$ to $\f$ for every type $T$ whose elements can be promoted.

Remember that, in our implementation of nonuniform coercions via unification
hints, we have assumed the ITP to only allow a coherent set of uniform
coercions. With our implementation, uniform coercions can now be handled as
special cases of nonuniform coercions, with the side effect of having the
possibility to declare incoherent sets. Incoherence is then handled for free
by systematically choosing the first matching hint (or the one with the
greatest priority).

\section{Reasoning about Curryfied morphisms}
\label{currymorph}
During the last year the Matita ITP was redesigned, and one of the novelties
was the introduction of unification hints to support 
the mathematical structures technique. Even if the old system (version 0.5.x)
was lacking any kind of structure inference support, we formalized with it a
hierarchy of algebraic and topological structures.  It was
clear that porting this formalization to the new system embracing
the mathematical structures approach was a good test bench. 
The following example, extracted from the formalization mentioned
above, convinced us to study the notion of nonuniform coercions 
presented in this paper.

The setting of the formalization is deeply extensional: points belongs
to types equipped with an equivalence relation, usually called setoids.
Functions respecting the equivalence relation of their source and target types
are called morphisms and are denoted with $A \Rightarrow B$ where
$A$ and $B$ are setoids: 
$$
\begin{array}{lcl}
\mathtt{setoid} & := & \{~ 
  T : \T;~ \approx : T \to T \to \P;~ 
  \pi_\approx : \mathtt{is\_equiv\_relation}~T~\approx ~\}
\\
A \Rightarrow B & := & \{~ 
  f : A \to B;~ 
  \pi_f : \forall x,y:A.x \approx_A y \to f~x \approx_B f~y ~\}
\end{array}
$$

One may expect that the concept of unary morphism 
extends in a straightforward way via currying to n-ary
morphisms as the concept of unary function extends, in an higher order
languages, to n-ary functions. This is not the case
for morphisms, since currying comes at an extra price: in order
to form $A \Rightarrow (B \Rightarrow C)$, the target $B \Rightarrow C$ must
be a setoid, while $B \Rightarrow C$ is a type. We can obtain a setoid of
carrier $B \Rightarrow C$ equipping the carrier with the following extensional
equality relation over morphisms:
$$
A,B : \mathtt{setoid};~ f,g:A \Rightarrow B \vdash f \approx_\Rightarrow g := 
  \forall x:A. f~x \approx_B g~x
$$
so that $(B \Rightarrow C, \approx_\Rightarrow B C, \ldots)$ is a setoid
(that we denote as $B \Rightarrow_\approx C$). It is now possible, but
cumbersome and distracting, to write a currified morphism as
$A \Rightarrow (B \Rightarrow_\approx C)$. A much better solution is to
exploit the following nonuniform coercion to allow the implicit promotion
of $B \Rightarrow C$ to $B \Rightarrow_\approx C$:
$$
 \begin{array}{lcl}
  A : \mathtt{setoid}, B : \mathtt{setoid} & \vdash &
    \begin{array}{lcl}
    \mathtt{Type} & \rightarrow & \mathtt{setoid} \\
    A\Rightarrow B & \mapsto & (A\Rightarrow B, \approx_\Rightarrow, \pi_\Rightarrow)
    \end{array}
 \end{array}
$$
\noindent
where $\pi_\Rightarrow$ is a 
proof that $\approx_\Rightarrow$ is an equivalence relation over morphisms.

Before developing the concept of non uniform coercions we tried other 
solutions, but they turned out to be quite unsatisfactory.

First, note that a uniform coercion cannot distinguish the type $A \Rightarrow
B$ from another type, thus would apply in unexpected contexts too. 

It would be also hard to achieve the same result relaying on the notational 
support the interactive theorem prover offers to overload the $\Rightarrow$
and $\Rightarrow_\approx$ notations. Indeed,
in the very same expression $A \Rightarrow (B \Rightarrow C)$
the first occurrence of $\Rightarrow$ is intended to define a morphism, while
the other occurrence must define a setoid, but interpreting all the
occurrences as setoids, possibly inserting a projection in front,
makes sense too (even if it is not the intended meaning).
This approach would thus require the system to employ a non trivial mechanism
to disambiguate notational overloading. 

Last, it is always possible to define different record types for unary, binary,
etc. morphisms, but it turns out to be rather inconvenient as soon as one has
to reason about partially instantiated morphisms. 
For example a partially instantiated binary morphism of type 
$A \Rightarrow B \Rightarrow C $ would have type $B \to C$, since
functions belonging to that morphism class have type
$A \to B \to C$ and not $A \to B \Rightarrow C$. The real shortcoming 
is not that the system has to enrich partially instantiated morphism,
but that the user has to link every n-ary morphism to $n-1$ possible
enrichments (and all of them comprise a different proof for the
$\pi_f$ component of the morphism structure).

\section{Lifting types, not notations}
\label{notation}

Every ITP allows some degree of user configurable notational support.
Consider the following conjecture (where we used the respectively infix 
notation $+$ for addition over integers and prefix notation $-$ for inverse)

$$x,y:\Z \vdash x + -(y + x) = -y$$
\noindent 
and the following well known fact: 
$$\mathtt{invmul}:\forall G:\G.\forall a,b:G.(a * b)^{-1} = b^{-1} * a^{-1}$$
\noindent
where $*$ is a notation for $(\op~G)$ and $\cdot^{-1}$ for $(\inv~G)$.
Thanks to unification hints, we can rewrite the conjecture using
$\mathtt{invmul}$ since the system knows that
$\Z$ together with $+$ and $-$ forms the \G{} $\GZ$. The result, however,
is quite confusing:
$$x,y:\Z \vdash x + x^{-1} * y^{-1} = -y$$
To understand it, it is better to de-activate the notation for group,
obtaining
$$x,y:\Z \vdash x + \op~\GZ~(\inv~\GZ~x)~(\inv~\GZ~y) = -y$$
The result is clearly correct, since $(\op~\GZ)$ reduces to $+$
and $(\inv~\GZ)$ to $-$, but, until we perform the reduction, the notation
we get is wrong.

This problem is already extremely frequent when we use the formalization
approach described in Section~\ref{mathcomp}, and it becomes worse
with nonuniform coercions, that may promote even an atomic term written by the
user (like $\zplus$, or $+$) to a richer operation, like multiplication in
a group (i.e. denoted with $*$).

A possible solution that avoids reducing the term is to overload the notation
for $+$ and $-$ over $(\op~\GZ)$ and $(\inv~\GZ)$. This is usually possible
since
the pattern $(\op~\GZ)$ is more precise than the pattern $(\op~\_)$ which is
matched in the generic group notation. Nevertheless, this is a bad solution
for two reasons: it requires to declare new notations every time we define
a new model of a given structure; and terms like $(\op~\GZ~x~y)$ are left in the
proof term where we would expect to find the simpler $x + y$.

A superior solution is forcing the needed reductions, without requiring any
user intervention. We achieve this in two steps. First of all, we change
the way we declare our hints so that the term that is left by unification
is no longer a projection of the form $(\op~\GZ)$, but a redex of the form
$$\op~\langle \Z, +, -, \mathtt{assoc}~\GZ, \mathtt{inv\_op\_cancel}~\GZ, \mathtt{unit\_law}~\GZ, \mathtt{closed\_op}~\GZ \rangle$$

Then we change the implementation
of our ITP so that redexes of this kind are automatically reduced as soon
as they are formed. In particular, if the projection in the redex retrieves the
carrier or an operation of the structure (like $\op$ or $\inv$), the reduction
yields the plain operation of the structure ($+$ or $-$); if the projection
retrieves a property, the reduction yields a projection, like
$\mathtt{assoc}~\GZ$ which is a compact proof term for the associativity
of addition.

The strategy of reducing this form of redexes
is consistent with the usually implemented one for 
$\beta$-redexes. For example, in a system like Coq or \MATITA, every
rewriting step with an equation $a=b$ is performed by first changing
the conjecture $P$ to the $\beta$-redex $((\lambda x.P[x/a]) a)$ and then
applying the elimination principle for equality. The rewritten conjecture
would be $((\lambda x.P[x/a]) b)$, but, since $\beta$-redex are immediately
reduced, we obtain $P[b/a]$.

The following is the hint able to generate the redexes in the case of the
group of integers. Compare it with the similar hint given
in Section~\ref{uhintsnut}:
\begin{prooftree}
\AxiomC{$?_g := \langle \Z, +, -, 0, 
         \mathtt{assoc}~\GZ, \mathtt{inv\_op\_cancel}~\GZ, 
         \mathtt{unit\_law}~\GZ, \mathtt{closed\_op}~\GZ \rangle $}
\LeftLabel{$\vdash$}
\UnaryInfC{ $ \op~?_g \equiv +$ }
\end{prooftree}

It is worth observing that the expanded form of the hinted solution can
obtained mechanically by separating fields that hold properties, that are kept
as projections of the $\GZ$ structure, from fields holding types or operations,
that on the contrary are expanded. 

It is natural to compare this approach with the alternative one consisting
in implementing an ad hoc simplification tactic that reduces only redexes 
involving projections applied to concrete instances of a structure. 
While this approach does not necessarily require any deep modification to 
the interactive theorem prover,\footnote{Assuming the ITP features a language
to let the user define new tactics, like $\mathcal{L}$-tac.} it has to be
triggered manually and its effect is not recorded in the proof term. Reduction
is not a proof step in CIC, thus the effect of any reduction tactic is left
implicit in the resulting proof term. Hardwiring the greedy reduction strategy
we propose in the proof engine allows to obtain proof term in which the
aforementioned redexes are not present. This affects in a positive way not only
the type checking time, thanks to the smaller size of proof terms, but also the
output quality of any procedure manipulating proof terms.\\ For example
in~\cite{fguidi-proc} and~\cite{csc-decl} the authors reconstruct a proof
script, respectively based on a procedural and a declarative tactic language,
starting from a proof term. Another example is~\cite{natural,YANNTHESIS} where
an explanation of a proof in natural language is obtained solely processing a
proof term. In all these cases a proof term were operations are of the form
$(\op~\GZ)$ would be clearly explained with the nomenclature of abstract group
theory, while the original proof was carried on the concrete setting of
integers, and the user did resort to abstract group theory only to justify some
of his proof steps.

\section{Conclusions}
\label{conclusions}

The most powerful tool of modern mathematics is abstraction in the following
sense: the mathematical corpus is no longer a flat collection of 
facts, but a hierarchy of algebraic, topological and geometrical theories, 
all characterized by its own set of axioms and often very rich in models. 
Every concrete mathematical object
belongs, directly or via isomorphisms, to several models and the mathematician
effortlessly mixes lemmas from all the relative theories.

This paper aims at making the formalization technique pioneered by
Gonthier et al.~\cite{BigOps,canonical-structures} and the general 
machinery behind it~\cite{unification-hints} run
smoothly, allowing the unification heuristic of ITPs to behave in a more
consistent way with respect to 
the inference of the mathematical structure a model
belongs to. In particular, the inference is not triggered only when 
projections of the structure are involved in a unification problem,
but more generally whenever a term (or type) has to be promoted to
a richer structure. The promotion is expressed in terms of nonuniform 
coercions, a novel notion to the authors knowledge, that are shown to 
be implementable in terms of uniform coercions and unification hints, 
in an higher order logic equipped with dependent types and pattern matching.



\paragraph{Acknowledgments} We would like to thank Jeremy Avigad for
proofreading preliminary versions of this paper.

\bibliographystyle{eptcs} 
\bibliography{../BIBTEX/helm}

\end{document}